\documentclass[3p]{elsarticle}

\usepackage{lineno,hyperref}
\usepackage{amsmath,amssymb}

\journal{Journal of Computational Physics}

\begin{document}

\begin{frontmatter}

\title{Quadratic conservative scheme for relativistic Vlasov--Maxwell system}


\author[tohoku]{Takashi Shiroto\corref{mycorrespondingauthor}}
\ead{tshiroto@rhd.mech.tohoku.ac.jp}
\cortext[mycorrespondingauthor]{Corresponding author}

\author[tohoku]{Naofumi Ohnishi}

\author[osaka]{Yasuhiko Sentoku}

\address[tohoku]{Department of Aerospace Engineering, Tohoku University, 6-6-01 Aramaki-Aza-Aoba, Aoba-ku, Sendai, Miyagi 980-8579, Japan}
\address[osaka]{Institute of Laser Engineering, Osaka University, 6-2 Yamadaoka, Suita, Osaka 565-0871, Japan}

\begin{abstract}
For more than half a century, most of the plasma scientists have encountered a violation of the conservation laws of
charge, momentum, and energy whenever they have numerically solve the first-principle equations of kinetic plasmas,
such as the relativistic Vlasov--Maxwell system.
This fatal problem is brought by the fact that both the Vlasov and Maxwell equations are indirectly associated
with the conservation laws by means of some mathematical manipulations.
Here we propose a quadratic conservative scheme, which can strictly maintain the conservation laws
by discretizing the relativistic Vlasov--Maxwell system.
A discrete product rule and summation-by-parts are the key players in the construction of the quadratic conservative scheme.
Numerical experiments of the relativistic two-stream instability and relativistic Weibel instability
prove the validity of our computational theory,
and the proposed strategy will open the doors to the first-principle studies of mesoscopic and macroscopic plasma physics.
\end{abstract}


\end{frontmatter}


\section{Introduction}\label{sec:1}
The relativistic Vlasov--Maxwell system has been regarded as the first-principle equations of weakly coupled plasmas.
The relativistic Vlasov equation is

\begin{align}
\frac{\partial f}{\partial t}+\frac{\partial}{\partial \mathbf{x}}\cdot\left(\frac{\mathbf{p}}{\gamma m}f\right)+
\frac{\partial}{\partial \mathbf{p}}\cdot\left[q\left(\mathbf{E}+\frac{\mathbf{p}\times\mathbf{B}}{\gamma mc}\right)f\right]=0,\label{eq:1.1}
\end{align}
where $f=f(t,\mathbf{x},\mathbf{p})$ is the distribution function; 
$t$, $\mathbf{x}=[x,y,z]^\mathrm{T}$, and $\mathbf{p}=[p_x,p_y,p_z]^\mathrm{T}$ are the time, space, and momentum, respectively;
$m$ and $q$ are the particle mass and charge, respectively;
$\mathbf{E}=[E_x,E_y,E_z]^\mathrm{T}$ and $\mathbf{B}=[B_x,B_y,B_z]^\mathrm{T}$ are the electric and magnetic field respectively;
and $c$ is the speed of light in vacuum. $\gamma$ is the Lorentz factor described as

\begin{align}
\gamma=\sqrt{1+\left|\frac{\mathbf{p}}{mc}\right|^2}.\label{eq:1.2}
\end{align}
This equation is coupled with the governing equations for the electromagnetic field: Maxwell's equations in Gaussian-cgs units:

\begin{align}
\mathrm{rot\ }\mathbf{B}=\frac{4\pi}{c}\mathbf{J}+\frac{1}{c}\frac{\partial \mathbf{E}}{\partial t},\label{eq:1.3}\\
\mathrm{rot\ }\mathbf{E}=-\frac{1}{c}\frac{\partial \mathbf{B}}{\partial t},\label{eq:1.4}\\
\mathrm{div\ }\mathbf{E}=4\pi \rho,\label{eq:1.5}\\
\mathrm{div\ }\mathbf{B}=0,\label{eq:1.6}
\end{align}
where $\rho$ and $\mathbf{J}$ are the charge and current densities, respectively.
However, Eqs.~(\ref{eq:1.5}) and (\ref{eq:1.6}) are naturally satisfied when the law of charge conservation
and the inexistence of the magnetic monopole are assumed.
Fortunately, these principles are derived from the 0th-order moment equation of Eq.~(\ref{eq:1.1}).
Therefore, we do not need to solve Eqs.~(\ref{eq:1.5}) and (\ref{eq:1.6}) when solving Eqs.~(\ref{eq:1.1}),
(\ref{eq:1.3}), and (\ref{eq:1.4}).

Modern numerical investigations of kinetic plasmas can be characterized in two ways.
The first one is a particle-in-cell (PIC) method \cite{Birdsall}, which solves the equations of motion of
charged particles, such as ions and electrons, instead of the Vlasov equation.
In the PIC, the equations of motion are coupled with the Maxwell equations using some of the field interpolation techniques.
Another approach is to discretize the Vlasov equations directly by using the finite-difference method, spectral method,
and so on (hereafter, called ``Vlasov simulation'').
However, these numerical methods have a fatal problem;
the conservation laws of charge, momentum, and energy are violated in principle when
the governing equations are discretized.
The term ``numerical heating'' is a nightmare among PIC users,
which implies that the total energies in PIC simulations increase infinitely even if there is no physical energy source.
To overcome this issue, many mathematical investigations were performed
on the conservation property of the first-principle kinetic simulations,
and significant progress was made mainly in the 2010s.
Crank--Nicolson time integration is one of the key structures in the construction of conservative PIC methods;
recent studies have employed it in energy-conserving \cite{Markidis2011,Cheng2014-1,Cheng2014-2},
charge-energy-conserving one-dimensional one-momentum-component (1D1P) \cite{Chen2011},
one-dimensional three-momentum-components (1D3P) \cite{Chen2014},
and two-dimensional three-momentum-component (2D3P) \cite{Chen2015} PIC methods.
A study discretized the equations of motion with a leap-frog method,
while the Maxwell equations were solved with the Crank--Nicolson method;
the total energy was strictly conserved with round-off errors,
but the charge conservation (Gauss's law) could not be maintained simultaneously \cite{Lapenta2017}.
It is difficult to construct a PIC method, which can strictly maintain the conservation laws of
charge, momentum, and energy because the distribution functions are solved by particles,
although the electromagnetic field is discretized by the finite-difference method.
In the particle methods, the shapes of the particles are strictly maintained after time integration;
that means particle methods are free from a numerical dispersion.
However, there are no dispersion-free finite-difference schemes.
This mathematical inconsistency makes it impossible to construct an exactly conservative PIC method.

Recently, exactly conservative Vlasov simulation schemes have been demonstrated for
Vlasov--Poission systems \cite{Camporeale2016,Manzini2016}, and Vlasov--Maxwell systems \cite{Delzanno2015}
using the Crank--Nicolson and spectral methods.
The spectral method \cite{Canuto} has no numerical dispersion; hence, it can overcome the problem of PIC methods.
In the numerical experiment, errors of the conservation laws were strictly maintained at the round-off level
when the tolerance of the Crank--Nicolson method was small enough.
However, the spectral method cannot employ non-periodic boundary conditions \cite{Mur1981},
so that these algorithms are applicable only to restricted situations.
To perform kinetic simulations with non-periodic boundaries,
a conservative Vlasov--Maxwell scheme based on the finite-difference manner is required.
In one work, the Vlasov equations of the conservative form were discretized with the
conservative form of the interpolated differential operator (IDO-CF) method \cite{Imai2008},
but the errors of energy conservation were much larger than the round-off level \cite{Imadera2009}.
In gyrokinetic simulations, there is a charge-conserving algorithm based on finite-difference methods,
although the momentum and energy cannot be conserved \cite{Idomura2007,Idomura2008,Idomura2016}.
In spite of their complex curvilinear coordinates, these codes employ the Morinishi scheme \cite{Morinishi1998}
to maintain the law of charge conservation strictly.

In the research field of computational fluid dynamics, the Morinishi scheme is regarded
as one of the quadratic conservative schemes.
The quadratic conservative schemes solve the time development of $f$ and $g$
and conserve the inner product $f\cdot g$ simultaneously.
According to these mathematical requirements, the quadratic conservative schemes
are based on some type of product rule in discrete form.
Although the original Morinishi scheme was composed for incompressible fluid dynamics,
this strategy has been extended to compressible fluid dynamics \cite{Morinishi2010},
and many other quadratic conservative schemes for compressible fluid dynamics have been proposed
\cite{Jameson2008,Subbareddy2009,Pirozzoli2010,Honein2004,Kuya2017}.
Further, the product rules in discrete form are useful in constructing conservative numerical methods
for hyperbolic hydrodynamic equations in non-conservative formulation \cite{Abgrall2003,Abgrall2017,Shiroto2017b}.
Such a strategy has been called a ``structure-preserving'' theory \cite{Morrison2017}.
A conservative algorithm should be constructed for the relativistic
Vlasov--Maxwell system using the structure-preserving strategy.

In this article, a quadratic conservative scheme is proposed for a relativistic Vlasov--Maxwell system,
which is based on the finite-difference method and strictly maintains the conservation laws of charge, momentum, and energy.
In Sec.~\ref{sec:2}, the quadratic conservative scheme for the relativistic Vlasov--Maxwell system is proposed.
The theoretical proof of Gauss's law, solenoidal constraint of the magnetic field, and conservation laws
of charge, momentum, and energy is given in Sec.~\ref{sec:3}.
Some mathematical formulae used in this study are also introduced.
Section~\ref{sec:4} describes the experimental demonstration of the conservation property
via the relativistic two-stream instability and relativistic Weibel instability.
Section~\ref{sec:5} gives the conclusions of this article.

\section{Structure-preserving theory for relativistic Vlasov--Maxwell system}\label{sec:2}
Before the quadratic conservative scheme for the relativistic Vlasov--Maxwell system is introduced,
some important strategies for constructing the proposed scheme are described here.
When proving the conservation laws of momentum and energy,
the product rule and integration-by-parts are required both in differential and discrete forms.
In addition, the commutative property of finite-difference operators is required
to derive Gauss's law and solenoidal constraint of the magnetic field.
Therefore, the finite-difference operators should have a linearity. Accordingly,
we employed the Crank--Nicolson method for the temporal difference, and a 2nd-order central difference method for
the spatial and momentum differences.

The quadratic conservative discretization of the relativistic Vlasov equation is

\begin{align}
\frac{\delta}{\delta t}[f^{n+\frac12,i_1,i_2,i_3,j_1,j_2,j_3}]\notag\\
+\frac{p_x^{j_1}}{\gamma^{\tilde{j_1},j_2,j_3}m}\frac{\delta}{\delta x}[f^{\hat{n},i_1,i_2,i_3,j_1,j_2,j_3}]
+\frac{p_y^{j_2}}{\gamma^{j_1,\tilde{j_2},j_3}m}\frac{\delta}{\delta y}[f^{\hat{n},i_1,i_2,i_3,j_1,j_2,j_3}]
+\frac{p_z^{j_3}}{\gamma^{j_1,j_2,\tilde{j_3}}m}\frac{\delta}{\delta z}[f^{\hat{n},i_1,i_2,i_3,j_1,j_2,j_3}]\notag\\
+qE_x^{\hat{n},i_1,i_2,i_3}\frac{\delta}{\delta p_x}[f^{\hat{n},i_1,i_2,i_3,j_1,j_2,j_3}]
+\frac{qB_z^{\hat{n},i_1,i_2,i_3}p_y^{j_2}}{mc}\frac{\delta}{\delta p_x}
\left[\frac{f^{\hat{n},i_1,i_2,i_3,j_1,j_2,j_3}}{\gamma^{j_1,\tilde{j_2},j_3}}\right]
-\frac{qB_y^{\hat{n},i_1,i_2,i_3}p_z^{j_3}}{mc}\frac{\delta}{\delta p_x}
\left[\frac{f^{\hat{n},i_1,i_2,i_3,j_1,j_2,j_3}}{\gamma^{j_1,j_2,\tilde{j_3}}}\right]\notag\\
+qE_y^{\hat{n},i_1,i_2,i_3}\frac{\delta}{\delta p_y}[f^{\hat{n},i_1,i_2,i_3,j_1,j_2,j_3}]
+\frac{qB_x^{\hat{n},i_1,i_2,i_3}p_z^{j_3}}{mc}\frac{\delta}{\delta p_y}
\left[\frac{f^{\hat{n},i_1,i_2,i_3,j_1,j_2,j_3}}{\gamma^{j_1,j_2,\tilde{j_3}}}\right]
-\frac{qB_z^{\hat{n},i_1,i_2,i_3}p_x^{j_1}}{mc}\frac{\delta}{\delta p_y}
\left[\frac{f^{\hat{n},i_1,i_2,i_3,j_1,j_2,j_3}}{\gamma^{\tilde{j_1},j_2,j_3}}\right]\notag\\
+qE_z^{\hat{n},i_1,i_2,i_3}\frac{\delta}{\delta p_z}[f^{\hat{n},i_1,i_2,i_3,j_1,j_2,j_3}]
+\frac{qB_y^{\hat{n},i_1,i_2,i_3}p_x^{j_1}}{mc}\frac{\delta}{\delta p_z}
\left[\frac{f^{\hat{n},i_1,i_2,i_3,j_1,j_2,j_3}}{\gamma^{\tilde{j_1},j_2,j_3}}\right]
-\frac{qB_x^{\hat{n},i_1,i_2,i_3}p_y^{j_2}}{mc}\frac{\delta}{\delta p_z}
\left[\frac{f^{\hat{n},i_1,i_2,i_3,j_1,j_2,j_3}}{\gamma^{j_1,\tilde{j_2},j_3}}\right]\notag\\
=0,\label{eq:2.1}
\end{align}
where $n$, $i_1\in[1,N_x]$, $i_2\in[1,N_y]$, $i_3\in[1,N_z]$, $j_1\in[1,M_x]$, $j_2\in[1,M_y]$, and $j_3\in[1,M_z]$ are the indices of
$t$, $x$, $y$, $z$, $p_x$, $p_y$, and $p_z$, respectively.
$\Delta t$, $\Delta x$, $\Delta y$, $\Delta z$, $\Delta p_x$, $\Delta p_y$, and $\Delta p_z$ are the grid intervals of
$t$, $x$, $y$, $z$, $p_x$, $p_y$, and $p_z$, respectively.
The finite-difference operators and interpolation operators are defined as follows:

\begin{align}
\frac{\delta}{\delta t}[F^{n+\frac12}]=\frac{F^{n+1}-F^n}{\Delta t},\
\frac{\delta}{\delta x}[F^{i_1}]=\frac{F^{i_1+1}-F^{i_1-1}}{\Delta x},\
\frac{\delta}{\delta y}[F^{i_2}]=\frac{F^{i_2+1}-F^{i_2-1}}{\Delta y},\
\frac{\delta}{\delta z}[F^{i_3}]=\frac{F^{i_3+1}-F^{i_3-1}}{\Delta z},\notag\\
\frac{\delta}{\delta p_x}[F^{j_1}]=\frac{F^{j_1+1}-F^{j_1-1}}{\Delta p_x},\
\frac{\delta}{\delta p_y}[F^{j_2}]=\frac{F^{j_2+1}-F^{j_2-1}}{\Delta p_y},\
\frac{\delta}{\delta p_z}[F^{j_3}]=\frac{F^{j_3+1}-F^{j_3-1}}{\Delta p_z},\notag\\
F^{\hat{n}}=\frac{F^{n+1}+F^n}{2},\
F^{\tilde{j_1}}=\frac{F^{j_1+1}+F^{j_1-1}}{2},\
F^{\tilde{j_2}}=\frac{F^{j_2+1}+F^{j_2-1}}{2},\
F^{\tilde{j_3}}=\frac{F^{j_3+1}+F^{j_3-1}}{2},\notag
\end{align}
where $F$ is an arbitrary function.
Moreover, the distribution function $f$ must be maintained at the round-off level near the momentum boundaries:

\begin{align}
f^{n,i_1,i_2,i_3,1,j_2,j_3} = f^{n,i_1,i_2,i_3,2,j_2,j_3} =
f^{n,i_1,i_2,i_3,M_x-1,j_2,j_3} = f^{n,i_1,i_2,i_3,M_x,j_2,j_3} = 0,\label{eq:2.2}\\
f^{n,i_1,i_2,i_3,j_1,1,j_3} = f^{n,i_1,i_2,i_3,j_1,2,j_3} =
f^{n,i_1,i_2,i_3,j_1,M_y-1,j_3} = f^{n,i_1,i_2,i_3,j_1,M_y,j_3} = 0,\label{eq:2.3}\\
f^{n,i_1,i_2,i_3,j_1,j_2,1} = f^{n,i_1,i_2,i_3,j_1,j_2,2} =
f^{n,i_1,i_2,i_3,j_1,j_2,M_z-1} = f^{n,i_1,i_2,i_3,j_1,j_2,M_z} = 0,\label{eq:2.4}
\end{align}
Therefore, the computational domain of momentum should be large enough to maintain Eqs.~(\ref{eq:2.2})--(\ref{eq:2.4})
Maxwell's equations (\ref{eq:1.3}) and (\ref{eq:1.4}) must be discretized as follows:

\begin{align}
\frac{\delta}{\delta y}[B_z^{\hat{n},i_1,i_2,i_3}]-\frac{\delta}{\delta z}[B_y^{\hat{n},i_1,i_2,i_3}]
=\frac{4\pi}{c}J_x^{\hat{n},i_1,i_2,i_3}+\frac{1}{c}\frac{\delta}{\delta t}[E_x^{n+\frac12,i_1,i_2,i_3}],\label{eq:2.5}\\
\frac{\delta}{\delta z}[B_x^{\hat{n},i_1,i_2,i_3}]-\frac{\delta}{\delta x}[B_z^{\hat{n},i_1,i_2,i_3}]
=\frac{4\pi}{c}J_y^{\hat{n},i_1,i_2,i_3}+\frac{1}{c}\frac{\delta}{\delta t}[E_y^{n+\frac12,i_1,i_2,i_3}],\label{eq:2.6}\\
\frac{\delta}{\delta x}[B_y^{\hat{n},i_1,i_2,i_3}]-\frac{\delta}{\delta y}[B_x^{\hat{n},i_1,i_2,i_3}]
=\frac{4\pi}{c}J_z^{\hat{n},i_1,i_2,i_3}+\frac{1}{c}\frac{\delta}{\delta t}[E_z^{n+\frac12,i_1,i_2,i_3}],\label{eq:2.7}\\
\frac{\delta}{\delta y}[E_z^{\hat{n},i_1,i_2,i_3}]-\frac{\delta}{\delta z}[E_y^{\hat{n},i_1,i_2,i_3}]
=-\frac{1}{c}\frac{\delta}{\delta t}[B_x^{n+\frac12,i_1,i_2,i_3}],\label{eq:2.8}\\
\frac{\delta}{\delta z}[E_x^{\hat{n},i_1,i_2,i_3}]-\frac{\delta}{\delta x}[E_z^{\hat{n},i_1,i_2,i_3}]
=-\frac{1}{c}\frac{\delta}{\delta t}[B_y^{n+\frac12,i_1,i_2,i_3}],\label{eq:2.9}\\
\frac{\delta}{\delta x}[E_y^{\hat{n},i_1,i_2,i_3}]-\frac{\delta}{\delta y}[E_x^{\hat{n},i_1,i_2,i_3}]
=-\frac{1}{c}\frac{\delta}{\delta t}[B_z^{n+\frac12,i_1,i_2,i_3}],\label{eq:2.10}
\end{align}
When solving Eqs.~(\ref{eq:2.5})--(\ref{eq:2.10}), the current density $\mathbf{J}$
should be obtained from the distribution function $f$.
Furthermore, Gauss's law Eq.~(\ref{eq:1.5}) is required to derive the law of momentum conservation,
which is associated with the charge density $\rho$.
Therefore, these quantities are defined as follows:

\begin{align}
\rho^{n,i_1,i_2,i_3}=q\sum_{j_1,j_2,j_3}
f^{n,i_1,i_2,i_3,j_1,j_2,j_3}\Delta V,\label{eq:2.11}\\
J^{n,i_1,i_2,i_3}_x=\frac{q}{m}\sum_{j_1,j_2,j_3}
\frac{p^{j_1}_x}{\gamma^{\tilde{j_1},j_2,j_3}} f^{n,i_1,i_2,i_3,j_1,j_2,j_3}\Delta V,\label{eq:2.12}\\
J^{n,i_1,i_2,i_3}_y=\frac{q}{m}\sum_{j_1,j_2,j_3}
\frac{p^{j_2}_y}{\gamma^{j_1,\tilde{j_2},j_3}} f^{n,i_1,i_2,i_3,j_1,j_2,j_3}\Delta V,\label{eq:2.13}\\
J^{n,i_1,i_2,i_3}_z=\frac{q}{m}\sum_{j_1,j_2,j_3}
\frac{p^{j_3}_z}{\gamma^{j_1,j_2,\tilde{j_3}}} f^{n,i_1,i_2,i_3,j_1,j_2,j_3}\Delta V,\label{eq:2.14}
\end{align}
where $\Delta V=\Delta p_x \Delta p_y \Delta p_z$.
Note that the domain of summation covers the entire computational domain:

\begin{align}
\sum_{j_1,j_2,j_3}=\sum_{j_1}\sum_{j_2}\sum_{j_3},\quad
\sum_{j_1}=\sum_{j_1=2}^{M_x-1},\quad
\sum_{j_2}=\sum_{j_2=2}^{M_y-1},\quad
\sum_{j_3}=\sum_{j_3=2}^{M_z-1}.\label{eq:2.15}
\end{align}
These constitute the complete set of our quadratic conservative scheme.

\section{Proof of conservation property}\label{sec:3}
This section gives the proof of the exact conservation properties of charge, momentum, and energy
for the proposed discretization method in the previous section.
First, the discrete product rule, summation-by-parts,
and commutative laws of finite-differential operators are derived in Sec.~\ref{sec:3.1}.
The law of charge conservation is derived in Sec.~\ref{sec:3.2}.
In Sec.~\ref{sec:3.3}, Gauss's law and the solenoidal constraint of the magnetic field are obtained
to prove the law of momentum conservation.
The laws of momentum and energy conservation are derived in Sec.~\ref{sec:3.4} and Sec.~\ref{sec:3.5}, respectively.

\subsection{Mathematical basis}\label{sec:3.1}
In this article, the following finite-difference operator is defined to prove the conservation laws of momentum and energy:

\begin{align}
\frac{\mathrm{D}}{\mathrm{D}x}[F^{i_1},G^{i_1}]\equiv
\frac{F^{i_1+1}G^{i_1}+F^{i_1}G^{i_1+1}-F^{i_1}G^{i_1-1}-F^{i_1-1}G^{i_1}}{2\Delta x},
\end{align}
where $F$ and $G$ are the arbitrary functions.
The equivalent operator has been used to construct kinetic-energy-preserving schemes \cite{Morinishi1998}.
A product rule for the momentum dimensions is defined as follows:

\begin{align}
\frac{\delta}{\delta p_x}[F^{j_1}]G^{j_1}+F^{j_1}\frac{\delta}{\delta p_x}[G^{j_1}]&=
\frac{F^{j_1+1}-F^{j_1-1}}{2\Delta p_x}G^{j_1}+F^{j_1}\frac{G^{j_1+1}-G^{j_1-1}}{2\Delta p_x}\notag\\
&=\frac{F^{j_1+1}G^{j_1}+F^{j_1}G^{j_1+1}-F^{j_1}G^{j_1-1}-F^{j_1-1}G^{j_1}}{2\Delta p_x}\notag\\
&=\frac{\mathrm{D}}{\mathrm{D} p_x}[F^{j_1},G^{j_1}].\label{eq:3.1.1}
\end{align}
Obviously, Eq.~(\ref{eq:3.1.1}) is also applicable to the spatial dimensions.
A formula of summation-by-parts is obtained from Eq.~(\ref{eq:3.1.1}):

\begin{align}
\sum_{j_1} \frac{\delta}{\delta p_x}[F^{j_1}] G^{j_1}\Delta p_x=
\frac{F^{M_x}G^{M_x-1}+F^{M_x-1}G^{M_x}-F^{2}G^{1}-F^{1}G^{2}}{2}
-\sum_{j_1}F^{j_1} \frac{\delta}{\delta p_x}[G^{j_1}] \Delta p_x.\label{eq:3.1.2}
\end{align}
If $F^{1}=F^{2}=F^{M_x-1}=F^{M_x}=0$ is assumed, Eq.~(\ref{eq:3.1.2}) can be written in a simpler form:

\begin{align}
\sum_{j_1} \frac{\delta}{\delta p_x}[F^{j_1}] G^{j_1}\Delta p_x=
-\sum_{j_1}F^{j_1} \frac{\delta}{\delta p_x}[G^{j_1}] \Delta p_x.\label{eq:3.1.3}
\end{align}
Consequently, the constraint Eqs.~(\ref{eq:2.2})--(\ref{eq:2.4}) are enforced.
Furthermore, another type of product rule is used for the time derivative:

\begin{align}
\frac{\delta}{\delta t}[F^{n+\frac12}]G^{\hat{n}}+F^{\hat{n}}\frac{\delta}{\delta t}[G^{n+\frac12}]
=\frac{\delta}{\delta t}[(FG)^{n+\frac12}].\label{eq:3.1.4}
\end{align}
This formula is used to obtain time derivatives of the momentum of the electromagnetic field $(\mathbf{E}\times\mathbf{B})/4\pi c$
and the energy of the electromagnetic field $(\mathbf{E}^2+\mathbf{B}^2)/8\pi$ from Eqs.~(\ref{eq:2.5})--(\ref{eq:2.10}).
To prove Gauss's law and the solenoidal constraint for the magnetic field,
the commutative laws of finite-difference operators are derived as follows:

\begin{align}
\frac{\delta}{\delta x}\left[\frac{\delta}{\delta y}[F^{i_1,i_2}]\right]
&=\frac{F^{i_1+1,j_2+1}-F^{i_1+1,j_2-1}-F^{i_1-1,i_2+1}+F^{i_1-1,i_2-1}}{\Delta x \Delta y}\notag\\
&=\frac{\delta}{\delta y}\left[\frac{\delta}{\delta x}[F^{i_1,i_2}]\right],\label{eq:3.1.5}\\
\frac{\delta}{\delta t}\left[\frac{\delta}{\delta x}[F^{n+\frac12,i_1}]\right]
&=\frac{\delta}{\delta x}\left[\frac{\delta}{\delta t}[F^{n+\frac12,i_1}]\right].\label{eq:3.1.6}
\end{align}
Corresponding to the proof of energy conservation, some formulae related to the Lorentz factor $\gamma$ are derived.
From the definition of $\gamma$;

\begin{align}
\frac{(\gamma^{j_1+1,j_2,j_3})^2-(\gamma^{j_1-1,j_2,j_3})^2}{2\Delta p_x}&=
\frac{1}{m^2c^2}\frac{(p^{j_1+1}_x)^2-(p^{j_1-1}_x)^2}{2\Delta p_x},\notag\\
\therefore\frac{\delta}{\delta p_x}[\gamma^{j_1,j_2,j_3}]
&=\frac{1}{m^2c^2}\frac{p^{j_1+1}_x+p^{j_1-1}_x}{\gamma^{j_1+1,j_2,j_3}+\gamma^{j_1-1,j_2,j_3}}
\frac{p^{j_1+1}_x-p^{j_1-1}_x}{2\Delta p_x}\notag\\
&=\frac{1}{m^2c^2}\frac{p^{j_1}_x}{\gamma^{\tilde{j_1},j_2,j_3}}
\neq\frac{1}{m^2c^2}\frac{p^{j_1}_x}{\gamma^{j_1,j_2,j_3}},\label{eq:3.1.7}\\
\frac{\delta}{\delta p_y}[\gamma^{j_1,j_2,j_3}]
&=\frac{1}{m^2c^2}\frac{p^{j_2}_y}{\gamma^{j_1,\tilde{j_2},j_3}},\label{eq:3.1.8}\\
\frac{\delta}{\delta p_z}[\gamma^{j_1,j_2,j_3}]
&=\frac{1}{m^2c^2}\frac{p^{j_3}_z}{\gamma^{j_1,j_2,\tilde{j_3}}}.\label{eq:3.1.9}
\end{align}

\subsection{The law of charge conservation}\label{sec:3.2}
The law of charge conservation is the 0th-order moment of the relativistic Vlasov equation.
In differential form, this is described as follows:

\begin{align}
q\frac{\partial}{\partial t}\int^{\infty}_{-\infty}\int^{\infty}_{-\infty}\int^{\infty}_{-\infty}
f \mathrm{d}p_x \mathrm{d}p_y \mathrm{d}p_z+
q\frac{\partial}{\partial \mathbf{x}}\cdot\int^{\infty}_{-\infty}\int^{\infty}_{-\infty}\int^{\infty}_{-\infty}
\frac{\mathbf{p}}{\gamma m}f \mathrm{d}p_x \mathrm{d}p_y \mathrm{d}p_z=0,\notag\\
\therefore \frac{\partial \rho}{\partial t}+\nabla\cdot\mathbf{J}=0.\label{eq:3.2.1}
\end{align}
The corresponding equation in discrete form is derived from Eq.~(\ref{eq:2.1}) as follows:

\begin{align}
\frac{\delta}{\delta t}\left[q\sum_{j_1,j_2,j_3}f^{n+\frac12,i_1,i_2,i_3,j_1,j_2,j_3}
\Delta V\right]
+\frac{\delta}{\delta x}\left[\frac{q}{m}\sum_{j_1,j_2,j_3}\frac{p^{j_1}_x}{\gamma^{\tilde{j_1},j_2,j_3}}
f^{\hat{n},i_1,i_2,i_3,j_1,j_2,j_3}\Delta V\right] \notag\\
+\frac{\delta}{\delta y}\left[\frac{q}{m}\sum_{j_1,j_2,j_3}\frac{p^{j_2}_y}{\gamma^{j_1,\tilde{j_2},j_3}}
f^{\hat{n},i_1,i_2,i_3,j_1,j_2,j_3}\Delta V\right]
+\frac{\delta}{\delta z}\left[\frac{q}{m}\sum_{j_1,j_2,j_3}\frac{p^{j_3}_z}{\gamma^{j_1,j_2,\tilde{j_3}}}
f^{\hat{n},i_1,i_2,i_3,j_1,j_2,j_3}\Delta V\right]=0.\label{eq:3.2.2}
\end{align}
By substituting Eqs.~(\ref{eq:2.11})--(\ref{eq:2.14}) into Eq.~(\ref{eq:3.2.2}), the following expression can be obtained:

\begin{align}
\frac{\delta}{\delta t}[\rho^{n+\frac12,i_1,i_2,i_3}]
+\frac{\delta}{\delta x}[J_x^{\hat{n},i_1,i_2,i_3}]
+\frac{\delta}{\delta y}[J_y^{\hat{n},i_1,i_2,i_3}]
+\frac{\delta}{\delta z}[J_z^{\hat{n},i_1,i_2,i_3}]=0.\label{eq:3.2.3}
\end{align}
Therefore, the law of charge conservation is strictly maintained, even in discrete form.
The conservation laws of particle number and mass are also derived similarly.

\subsection{Gauss's law and solenoidal constraint}\label{sec:3.3}
Here we review the derivation of Gauss's law in differential form.
The divergence of Eq.~(\ref{eq:1.3}) is

\begin{align}
\mathrm{div\ rot\ }\mathbf{B}=\frac{4\pi}{c}\mathrm{div\ }\mathbf{J}+\frac{1}{c}\mathrm{div\ }\frac{\partial \mathbf{E}}{\partial t},\notag\\
\therefore\frac{1}{c}\frac{\partial}{\partial t}(\mathrm{div\ }\mathbf{E})+\frac{4\pi}{c}\mathrm{div\ }\mathbf{J}=0.\label{eq:3.3.1}
\end{align}
Substituting Eq.~(\ref{eq:3.2.1}) into Eq.~(\ref{eq:3.3.1});

\begin{align}
\frac{1}{c}\frac{\partial}{\partial t}(\mathrm{div\ }\mathbf{E}-4\pi\rho)=0.\label{eq:3.3.6}
\end{align}
Thus, Gauss's law is maintained if the following condition is satisfied at the initial state:

\begin{align}
\mathrm{div\ }\mathbf{E}=4\pi\rho.\label{eq:3.3.7}
\end{align}

To reproduce these operations in discrete form, Eqs.~(\ref{eq:2.5})--(\ref{eq:2.7}) are transformed into the following form
using the commutative laws of finite-difference operators Eqs.~(\ref{eq:3.1.5}) and (\ref{eq:3.1.6}):

\begin{align}
\frac{\delta}{\delta x}\left[\frac{\delta}{\delta y}[B_z^{\hat{n},i_1,i_2,i_3}]\right]
-\frac{\delta}{\delta x}\left[\frac{\delta}{\delta z}[B_y^{\hat{n},i_1,i_2,i_3}]\right]
=\frac{4\pi}{c}\frac{\delta}{\delta x}[J_x^{\hat{n},i_1,i_2,i_3}]
+\frac{1}{c}\frac{\delta}{\delta x}\left[\frac{\delta}{\delta t}[E_x^{n+\frac12,i_1,i_2,i_3}]\right],\notag\\
\frac{\delta}{\delta y}\left[\frac{\delta}{\delta z}[B_x^{\hat{n},i_1,i_2,i_3}]\right]
-\frac{\delta}{\delta y}\left[\frac{\delta}{\delta x}[B_z^{\hat{n},i_1,i_2,i_3}]\right]
=\frac{4\pi}{c}\frac{\delta}{\delta y}[J_y^{\hat{n},i_1,i_2,i_3}]
+\frac{1}{c}\frac{\delta}{\delta y}\left[\frac{\delta}{\delta t}[E_y^{n+\frac12,i_1,i_2,i_3}]\right],\notag\\
\frac{\delta}{\delta z}\left[\frac{\delta}{\delta x}[B_y^{\hat{n},i_1,i_2,i_3}]\right]
-\frac{\delta}{\delta z}\left[\frac{\delta}{\delta y}[B_x^{\hat{n},i_1,i_2,i_3}]\right]
=\frac{4\pi}{c}\frac{\delta}{\delta z}[J_z^{\hat{n},i_1,i_2,i_3}]
+\frac{1}{c}\frac{\delta}{\delta z}\left[\frac{\delta}{\delta t}[E_z^{n+\frac12,i_1,i_2,i_3}]\right],\notag\\
\therefore \frac{1}{c}\frac{\delta}{\delta t}\left[
\frac{\delta}{\delta x}[E_x^{n+\frac12,i_1,i_2,i_3}]
+\frac{\delta}{\delta y}[E_y^{n+\frac12,i_1,i_2,i_3}]
+\frac{\delta}{\delta z}[E_z^{n+\frac12,i_1,i_2,i_3}]\right]\notag\\
+\frac{4\pi}{c}\left(
\frac{\delta}{\delta x}[J_x^{\hat{n},i_1,i_2,i_3}]
+\frac{\delta}{\delta y}[J_y^{\hat{n},i_1,i_2,i_3}]
+\frac{\delta}{\delta z}[J_z^{\hat{n},i_1,i_2,i_3}]
\right)=0.\label{eq:3.3.5}
\end{align}
Therefore, Eq.~(\ref{eq:3.3.1}) is automatically maintained by the above discretization.
By substituting the law of charge conservation Eq.~(\ref{eq:3.2.3}) into Eq.~(\ref{eq:3.3.5}),
a recurrence formula can be obtained:

\begin{align}
\frac{1}{c}\frac{\delta}{\delta t}\left[
\frac{\delta}{\delta x}[E_x^{n+\frac12,i_1,i_2,i_3}]
+\frac{\delta}{\delta y}[E_y^{n+\frac12,i_1,i_2,i_3}]
+\frac{\delta}{\delta z}[E_z^{n+\frac12,i_1,i_2,i_3}]
-4\pi\rho^{n+\frac12,i_1,i_2,i_3}\right]=0,\notag\\
\therefore
\frac{\delta}{\delta x}[E_x^{n+1,i_1,i_2,i_3}]
+\frac{\delta}{\delta y}[E_y^{n+1,i_1,i_2,i_3}]
+\frac{\delta}{\delta z}[E_z^{n+1,i_1,i_2,i_3}]
-4\pi\rho^{n+1,i_1,i_2,i_3}\notag\\
=\frac{\delta}{\delta x}[E_x^{n,i_1,i_2,i_3}]
+\frac{\delta}{\delta y}[E_y^{n,i_1,i_2,i_3}]
+\frac{\delta}{\delta z}[E_z^{n,i_1,i_2,i_3}]
-4\pi\rho^{n,i_1,i_2,i_3}\notag\\
\vdots\notag\\
=\frac{\delta}{\delta x}[E_x^{0,i_1,i_2,i_3}]
+\frac{\delta}{\delta y}[E_y^{0,i_1,i_2,i_3}]
+\frac{\delta}{\delta z}[E_z^{0,i_1,i_2,i_3}]
-4\pi\rho^{0,i_1,i_2,i_3}.
\end{align}
Hence, Gauss's law is strictly maintained even in discrete form
if the law is satisfied at the initial state:

\begin{align}
\frac{\delta}{\delta x}[E_x^{n+1,i_1,i_2,i_3}]
+\frac{\delta}{\delta y}[E_y^{n+1,i_1,i_2,i_3}]
+\frac{\delta}{\delta z}[E_z^{n+1,i_1,i_2,i_3}]
=4\pi\rho^{n+1,i_1,i_2,i_3}\notag\\
\mathrm{if}\quad
\frac{\delta}{\delta x}[E_x^{0,i_1,i_2,i_3}]
+\frac{\delta}{\delta y}[E_y^{0,i_1,i_2,i_3}]
+\frac{\delta}{\delta z}[E_z^{0,i_1,i_2,i_3}]
=4\pi\rho^{0,i_1,i_2,i_3}.\label{eq:3.3.9}
\end{align}

Likewise, the solenoidal constraint of the magnetic field is strictly maintained, even in discrete form:

\begin{align}
\frac{\delta}{\delta x}[B_x^{n+1,i_1,i_2,i_3}]
+\frac{\delta}{\delta y}[B_y^{n+1,i_1,i_2,i_3}]
+\frac{\delta}{\delta z}[B_z^{n+1,i_1,i_2,i_3}]
=0\notag\\
\mathrm{if}\quad
\frac{\delta}{\delta x}[B_x^{0,i_1,i_2,i_3}]
+\frac{\delta}{\delta y}[B_y^{0,i_1,i_2,i_3}]
+\frac{\delta}{\delta z}[B_z^{0,i_1,i_2,i_3}]
=0.\label{eq:3.3.10}
\end{align}

\subsection{The law of momentum conservation}\label{sec:3.4}
In this section, only the law of momentum conservation in the $x$-direction is discussed.
Here we review the derivation of momentum conservation in differential form.
The momentum of particles is described by the 1st-order moment of the relativistic Vlasov equation:

\begin{align}
\frac{\partial}{\partial t}\int^{\infty}_{-\infty}\int^{\infty}_{-\infty}\int^{\infty}_{-\infty}
p_x f \mathrm{d}p_x \mathrm{d}p_y \mathrm{d}p_z+
\frac{\partial}{\partial \mathbf{x}}\cdot\int^{\infty}_{-\infty}\int^{\infty}_{-\infty}\int^{\infty}_{-\infty}
p_x \frac{\mathbf{p}}{\gamma m} f \mathrm{d}p_x \mathrm{d}p_y \mathrm{d}p_z\notag\\
=-\int^{\infty}_{-\infty}\int^{\infty}_{-\infty}\int^{\infty}_{-\infty}
p_x \frac{\partial}{\partial \mathbf{p}}\cdot\left\{q\left(
\mathbf{E}+\frac{\mathbf{p}\times\mathbf{B}}{\gamma mc}\right)f\right\}
\mathrm{d}p_x \mathrm{d}p_y \mathrm{d}p_z\notag\\
=\int^{\infty}_{-\infty}\int^{\infty}_{-\infty}\int^{\infty}_{-\infty}
\frac{\partial p_x}{\partial \mathbf{p}} \cdot\left\{q\left(
\mathbf{E}+\frac{\mathbf{p}\times\mathbf{B}}{\gamma mc}\right)f\right\}
\mathrm{d}p_x \mathrm{d}p_y \mathrm{d}p_z\notag\\
=\rho E_x+\frac{J_y B_z-J_z B_y}{c}.\label{eq:3.4.1}
\end{align}
The Maxwell's equations are transformed into the momentum of the electromagnetic field with the product rule:

\begin{align}
\frac{B_z}{4\pi}\left(\frac{1}{c}\frac{\partial E_y}{\partial t}+\frac{\partial B_z}{\partial x}-\frac{\partial B_x}{\partial z}\right)
+\frac{E_y}{4\pi}\left(\frac{1}{c}\frac{\partial B_z}{\partial t}+\frac{\partial E_y}{\partial x}-\frac{\partial E_x}{\partial y}\right)
=-\frac{J_y B_z}{c},\notag\\
-\frac{B_y}{4\pi}\left(\frac{1}{c}\frac{\partial E_z}{\partial t}+\frac{\partial B_x}{\partial y}-\frac{\partial B_y}{\partial x}\right)
-\frac{E_z}{4\pi}\left(\frac{1}{c}\frac{\partial B_y}{\partial t}+\frac{\partial E_x}{\partial z}-\frac{\partial E_z}{\partial x}\right)
=\frac{J_z B_y}{c},\notag\\
\therefore \frac{1}{4\pi c}\frac{\partial (E_y B_z-E_z B_y)}{\partial t}
+\frac{1}{8\pi}\frac{\partial (|\mathbf{E}|^2+|\mathbf{B}|^2)}{\partial x}
-\frac{1}{4\pi}\mathrm{div\ }(E_x\mathbf{E}+B_x\mathbf{B})\notag\\
=-\frac{J_y B_z-J_z B_y}{c}
-\frac{E_x}{4\pi}\mathrm{div\ }\mathbf{E}-\frac{B_x}{4\pi}\mathrm{div\ }\mathbf{B}\notag\\
=-\rho E_x-\frac{J_y B_z-J_z B_y}{c}.\label{eq:3.4.10}
\end{align}
Finally, the law of momentum conservation is obtained because the terms on right-hand-side of
Eqs.~(\ref{eq:3.4.1}) and (\ref{eq:3.4.10}) cancel out:

\begin{align}
\frac{\partial}{\partial t}\left(\int^{\infty}_{-\infty}\int^{\infty}_{-\infty}\int^{\infty}_{-\infty}
p_x f \mathrm{d}p_x \mathrm{d}p_y \mathrm{d}p_z+
\frac{E_y B_z-E_z B_y}{4\pi c}\right)\notag\\
+\frac{\partial}{\partial x}\left(\int^{\infty}_{-\infty}\int^{\infty}_{-\infty}\int^{\infty}_{-\infty}
p_x \frac{p_x}{\gamma m} f \mathrm{d}p_x \mathrm{d}p_y \mathrm{d}p_z
+\frac{-{E_x}^2+{E_y}^2+{E_z}^2-{B_x}^2+{B_y}^2+{B_z}^2}{8\pi}\right)\notag\\
+\frac{\partial}{\partial y}\left(\int^{\infty}_{-\infty}\int^{\infty}_{-\infty}\int^{\infty}_{-\infty}
p_x \frac{p_y}{\gamma m} f \mathrm{d}p_x \mathrm{d}p_y \mathrm{d}p_z
-\frac{E_x E_y+B_x B_y}{4\pi}\right)\notag\\
+\frac{\partial}{\partial z}\left(\int^{\infty}_{-\infty}\int^{\infty}_{-\infty}\int^{\infty}_{-\infty}
p_x \frac{p_z}{\gamma m} f \mathrm{d}p_x \mathrm{d}p_y \mathrm{d}p_z
-\frac{E_x E_z+B_x B_z}{4\pi}\right)=0.\label{eq:3.4.12}
\end{align}

To obtain this relationship in discrete form, the following equations are derived by the summation-by-parts of Eq.~(\ref{eq:3.1.3}):

\begin{align}
\sum_{j_1,j_2,j_3}p^{j_1}_x
\frac{\delta}{\delta p_x}[f^{\hat{n},i_1,i_2,i_3,j_1,j_2,j_3}]\Delta V
=-\sum_{j_1,j_2,j_3} f^{\hat{n},i_1,i_2,i_3,j_1,j_2,j_3} \Delta V, \label{eq:3.4.2}\\
\sum_{j_1,j_2,j_3} p^{j_1}_x p^{j_2}_y
\frac{\delta}{\delta p_x}\left[\frac{f^{\hat{n},i_1,i_2,i_3,j_1,j_2,j_3}}{\gamma^{j_1,\tilde{j_2},j_3}}\right]\Delta V
=-\sum_{j_1,j_2,j_3} \frac{p^{j_2}_y}{\gamma^{j_1,\tilde{j_2},j_3}} f^{\hat{n},i_1,i_2,i_3,j_1,j_2,j_3}\Delta V,\label{eq:3.4.3}\\
\sum_{j_1,j_2,j_3} p^{j_1}_x p^{j_3}_z
\frac{\delta}{\delta p_x}\left[\frac{f^{\hat{n},i_1,i_2,i_3,j_1,j_2,j_3}}{\gamma^{j_1,j_2,\tilde{j_3}}}\right]\Delta V
=-\sum_{j_1,j_2,j_3} \frac{p^{j_3}_z}{\gamma^{j_1,j_2,\tilde{j_3}}} f^{\hat{n},i_1,i_2,i_3,j_1,j_2,j_3}\Delta V.\label{eq:3.4.4}
\end{align}
Thus, the momentum of particles in discrete form is described using Eqs.~(\ref{eq:3.4.2})--(\ref{eq:3.4.4}) as follows:

\begin{align}
\frac{\delta}{\delta t}\left[\sum_{j_1,j_2,j_3}p_x^{j_1}f^{n+\frac12,i_1,i_2,i_3,j_1,j_2,j_3}\Delta V\right]
+\frac{\delta}{\delta x}\left[\sum_{j_1,j_2,j_3}\frac{p_x^{j_1}p_x^{j_1}}{\gamma^{\tilde{j_1},j_2,j_3}m}
f^{\hat{n},i_1,i_2,i_3,j_1,j_2,j_3}\Delta V\right]\notag\\
+\frac{\delta}{\delta y}\left[\sum_{j_1,j_2,j_3}\frac{p_x^{j_1}p_y^{j_2}}{\gamma^{j_1,\tilde{j_2},j_3}m}
f^{\hat{n},i_1,i_2,i_3,j_1,j_2,j_3}\Delta V\right]
+\frac{\delta}{\delta z}\left[\sum_{j_1,j_2,j_3}\frac{p_x^{j_1}p_z^{j_3}}{\gamma^{j_1,j_2,\tilde{j_3}}m}
f^{\hat{n},i_1,i_2,i_3,j_1,j_2,j_3}\Delta V\right]\notag\\
=\sum_{j_1,j_2,j_3}\left(qE^{\hat{n},i_1,i_2,i_3}_x+\frac{qp_y^{j_2}B_z^{\hat{n},i_1,i_2,i_3}}{\gamma^{j_1,\tilde{j_2},j_3}mc}
-\frac{qp_z^{j_3}B_y^{\hat{n},i_1,i_2,i_3}}{\gamma^{j_1,j_2,\tilde{j_3}}mc} \right)f^{\hat{n},i_1,i_2,i_3,j_1,j_2,j_3}\Delta V\notag\\
=\rho^{\hat{n},i_1,i_2,i_3}E^{\hat{n},i_1,i_2,i_3}_x+
\frac{J^{\hat{n},i_1,i_2,i_3}_y B^{\hat{n},i_1,i_2,i_3}_z-J^{\hat{n},i_1,i_2,i_3}_z B^{\hat{n},i_1,i_2,i_3}_y}{c}.\label{eq:3.4.5}
\end{align}
To obtain the momentum of the electromagnetic field in discrete form, the temporal product rule Eq.~(\ref{eq:3.1.4}),
and the spatial product rule Eq.~(\ref{eq:3.1.1}) are applied to
Eqs.~(\ref{eq:2.6}), (\ref{eq:2.7}), (\ref{eq:2.9}), and (\ref{eq:2.10}):

\begin{align}
\frac{B_z^{\hat{n},i_1,i_2,i_3}}{4\pi}\left( \frac{1}{c}\frac{\delta}{\delta t}[E_y^{n+\frac12,i_1,i_2,i_3}]
+\frac{\delta}{\delta x}[B_z^{\hat{n},i_1,i_2,i_3}]-\frac{\delta}{\delta z}[B_x^{\hat{n},i_1,i_2,i_3}]\right)
=-\frac{1}{c}J_y^{\hat{n},i_1,i_2,i_3}B_z^{\hat{n},i_1,i_2,i_3},\notag\\
-\frac{B_y^{\hat{n},i_1,i_2,i_3}}{4\pi}\left( \frac{1}{c}\frac{\delta}{\delta t}[E_z^{n+\frac12,i_1,i_2,i_3}]
+\frac{\delta}{\delta y}[B_x^{\hat{n},i_1,i_2,i_3}]-\frac{\delta}{\delta x}[B_y^{\hat{n},i_1,i_2,i_3}]\right)
=\frac{1}{c}J_z^{\hat{n},i_1,i_2,i_3}B_y^{\hat{n},i_1,i_2,i_3},\notag\\
-\frac{E_z^{\hat{n},i_1,i_2,i_3}}{4\pi}\left( \frac{1}{c}\frac{\delta}{\delta t}[B_y^{n+\frac12,i_1,i_2,i_3}]
-\frac{\delta}{\delta x}[E_z^{\hat{n},i_1,i_2,i_3}]+\frac{\delta}{\delta z}[E_x^{\hat{n},i_1,i_2,i_3}]\right)=0,\notag\\
\frac{E_y^{\hat{n},i_1,i_2,i_3}}{4\pi}\left( \frac{1}{c}\frac{\delta}{\delta t}[B_z^{n+\frac12,i_1,i_2,i_3}]
-\frac{\delta}{\delta y}[E_x^{\hat{n},i_1,i_2,i_3}]+\frac{\delta}{\delta x}[E_y^{\hat{n},i_1,i_2,i_3}]\right)=0,\notag\\
\therefore \frac{\delta}{\delta t}\left[\frac{(E_yB_z-E_zB_y)^{n+\frac12,i_1,i_2,i_3}}{4\pi c}\right]
-\frac{1}{8\pi}\frac{\mathrm{D}}{\mathrm{D} x}[E_x^{\hat{n},i_1,i_2,i_3},E_x^{\hat{n},i_1,i_2,i_3}]
+\frac{1}{8\pi}\frac{\mathrm{D}}{\mathrm{D} x}[E_y^{\hat{n},i_1,i_2,i_3},E_y^{\hat{n},i_1,i_2,i_3}]\notag\\
+\frac{1}{8\pi}\frac{\mathrm{D}}{\mathrm{D} x}[E_z^{\hat{n},i_1,i_2,i_3},E_z^{\hat{n},i_1,i_2,i_3}]
-\frac{1}{8\pi}\frac{\mathrm{D}}{\mathrm{D} x}[B_x^{\hat{n},i_1,i_2,i_3},B_x^{\hat{n},i_1,i_2,i_3}]
+\frac{1}{8\pi}\frac{\mathrm{D}}{\mathrm{D} x}[B_y^{\hat{n},i_1,i_2,i_3},B_y^{\hat{n},i_1,i_2,i_3}]\notag\\
+\frac{1}{8\pi}\frac{\mathrm{D}}{\mathrm{D} x}[B_z^{\hat{n},i_1,i_2,i_3},B_z^{\hat{n},i_1,i_2,i_3}]
-\frac{1}{4\pi}\frac{\mathrm{D}}{\mathrm{D} y}[E_x^{\hat{n},i_1,i_2,i_3},E_y^{\hat{n},i_1,i_2,i_3}]
-\frac{1}{4\pi}\frac{\mathrm{D}}{\mathrm{D} y}[B_x^{\hat{n},i_1,i_2,i_3},B_y^{\hat{n},i_1,i_2,i_3}]\notag\\
-\frac{1}{4\pi}\frac{\mathrm{D}}{\mathrm{D} z}[E_x^{\hat{n},i_1,i_2,i_3},E_z^{\hat{n},i_1,i_2,i_3}]
-\frac{1}{4\pi}\frac{\mathrm{D}}{\mathrm{D} z}[B_x^{\hat{n},i_1,i_2,i_3},B_z^{\hat{n},i_1,i_2,i_3}]\notag\\
=-\rho^{\hat{n},i_1,i_2,i_3}E^{\hat{n},i_1,i_2,i_3}_x-
\frac{J^{\hat{n},i_1,i_2,i_3}_y B^{\hat{n},i_1,i_2,i_3}_z-J^{\hat{n},i_1,i_2,i_3}_z B^{\hat{n},i_1,i_2,i_3}_y}{c},\label{eq:3.4.6}
\end{align}
where the Gauss's law Eq.~(\ref{eq:3.3.9}) and the solenoidal constraint Eq.~(\ref{eq:3.3.10}) are used to derive the right-hand-side.
Therefore, the total momentum of charged particles and electromagnetic field is strictly conserved, even in discrete form
because the terms on the right-hand-side of Eqs.~(\ref{eq:3.4.5}) and (\ref{eq:3.4.6}) completely cancel out.
Although the proof is omitted, the law of momentum conservation is also derived for the $y$-direction and $z$-direction,
even in discrete form.

\subsection{The law of energy conservation}\label{sec:3.5}
Here we review the derivation of energy conservation in differential form.
The energy of particles is described by the 2nd-order moment of the relativistic Vlasov equation:

\begin{align}
\frac{\partial}{\partial t} \int^{\infty}_{-\infty}\int^{\infty}_{-\infty}\int^{\infty}_{-\infty}
\gamma mc^2f \mathrm{d}p_x \mathrm{d}p_y \mathrm{d}p_z
+\frac{\partial}{\partial \mathbf{x}}  \cdot\int^{\infty}_{-\infty}\int^{\infty}_{-\infty}\int^{\infty}_{-\infty}
c^2\mathbf{p}f \mathrm{d}p_x \mathrm{d}p_y \mathrm{d}p_z\notag\\
=-\int^{\infty}_{-\infty}\int^{\infty}_{-\infty}\int^{\infty}_{-\infty}
\gamma mc^2 \frac{\partial}{\partial \mathbf{p}}\cdot\left\{q\left(
\mathbf{E}+\frac{\mathbf{p}\times\mathbf{B}}{\gamma mc}\right)f\right\}
\mathrm{d}p_x \mathrm{d}p_y \mathrm{d}p_z\notag\\
=\int^{\infty}_{-\infty}\int^{\infty}_{-\infty}\int^{\infty}_{-\infty}
mc^2\frac{\partial \gamma}{\partial \mathbf{p}} \cdot\left\{q\left(
\mathbf{E}+\frac{\mathbf{p}\times\mathbf{B}}{\gamma mc}\right)f\right\}
\mathrm{d}p_x \mathrm{d}p_y \mathrm{d}p_z\notag\\
=\int^{\infty}_{-\infty}\int^{\infty}_{-\infty}\int^{\infty}_{-\infty}
\frac{\mathbf{p}}{\gamma m} \cdot\left\{q\left(
\mathbf{E}+\frac{\mathbf{p}\times\mathbf{B}}{\gamma mc}\right)f\right\}
\mathrm{d}p_x \mathrm{d}p_y \mathrm{d}p_z\notag\\
=\mathbf{J}\cdot\mathbf{E}.\label{eq:3.5.1}
\end{align}
The Amp\`{e}re--Maxwell and Faraday--Maxwell equations are transformed into the energy of the electromagnetic field by the product rule:

\begin{align}
\frac{c}{4\pi}E_x\left(
\frac{1}{c}\frac{\partial E_x}{\partial x}-\frac{\partial B_z}{\partial y}+\frac{\partial B_y}{\partial z}\right)
+\frac{c}{4\pi}E_y\left(
\frac{1}{c}\frac{\partial E_y}{\partial x}-\frac{\partial B_x}{\partial z}+\frac{\partial B_z}{\partial x}\right)
+\frac{c}{4\pi}E_z\left(
\frac{1}{c}\frac{\partial E_z}{\partial x}-\frac{\partial B_y}{\partial x}+\frac{\partial B_x}{\partial y}\right)
=-\mathbf{J}\cdot\mathbf{E},\notag\\
\frac{c}{4\pi}B_x\left(
\frac{1}{c}\frac{\partial B_x}{\partial x}+\frac{\partial E_z}{\partial y}-\frac{\partial E_y}{\partial z}\right)
+\frac{c}{4\pi}B_y\left(
\frac{1}{c}\frac{\partial B_y}{\partial x}+\frac{\partial E_x}{\partial z}-\frac{\partial E_z}{\partial x}\right)
+\frac{c}{4\pi}B_z\left(
\frac{1}{c}\frac{\partial B_z}{\partial x}+\frac{\partial E_y}{\partial x}-\frac{\partial E_x}{\partial y}\right)
=0,\notag\\
\therefore \frac{1}{8\pi}\frac{\partial (\mathbf{E}^2+\mathbf{B}^2)}{\partial t}+
\frac{c}{4\pi}\mathrm{div\ }(\mathbf{E}\times\mathbf{B})=-\mathbf{J}\cdot\mathbf{E}.\label{eq:3.5.10}
\end{align}
Finally, the law of energy conservation is obtained because the terms on the right-hand-side of Eqs.~(\ref{eq:3.5.1})
and (\ref{eq:3.5.10}) cancel out:

\begin{align}
\frac{\partial}{\partial t} \left(\int^{\infty}_{-\infty}\int^{\infty}_{-\infty}\int^{\infty}_{-\infty}
\gamma mc^2f \mathrm{d}p_x \mathrm{d}p_y \mathrm{d}p_z
+\frac{\mathbf{E}^2+\mathbf{B}^2}{8\pi}\right)
+\mathrm{div\ }  \left(\int^{\infty}_{-\infty}\int^{\infty}_{-\infty}\int^{\infty}_{-\infty}
c^2 \mathbf{p} f \mathrm{d}p_x \mathrm{d}p_y \mathrm{d}p_z
+c\frac{\mathbf{E}\times\mathbf{B}}{4\pi}\right)=0.\label{eq:3.5.12}
\end{align}

To obtain this relationship in discrete form,
the following relationships are derived by the summation-by-parts of Eq.~(\ref{eq:3.1.3}):

\begin{align}
\sum_{j_1,j_2,j_3} \gamma^{j_1,j_2,j_3} p^{j_3}_z \frac{\delta}{\delta p_y}
\left[\frac{f^{\hat{n},i_1,i_2,i_3,j_1,j_2,j_3}}{\gamma^{j_1,j_2,\tilde{j_3}}}\right]\Delta V
-\sum_{j_1,j_2,j_3} \gamma^{j_1,j_2,j_3} p^{j_2}_y \frac{\delta}{\delta p_z}
\left[\frac{f^{\hat{n},i_1,i_2,i_3,j_1,j_2,j_3}}{\gamma^{j_1,\tilde{j_2},j_3}}\right]\Delta V\notag\\
=\sum_{j_1,j_2,j_3} \left(-\frac{\delta}{\delta p_y}[\gamma^{j_1,j_2,j_3}]
\frac{p^{j_3}_z}{\gamma^{j_1,j_2,\tilde{j_3}}}
+\frac{\delta}{\delta p_z}[\gamma^{j_1,j_2,j_3}]
\frac{p^{j_2}_y}{\gamma^{j_1,\tilde{j_2},j_3}}\right) f^{\hat{n},i_1,i_2,i_3,j_1,j_2,j_3}\Delta V=0,\label{eq:3.5.2}\\
\sum_{j_1,j_2,j_3} \gamma^{j_1,j_2,j_3} p^{j_1}_x \frac{\delta}{\delta p_z}
\left[\frac{f^{\hat{n},i_1,i_2,i_3,j_1,j_2,j_3}}{\gamma^{\tilde{j_1},j_2,j_3}}\right]\Delta V
-\sum_{j_1,j_2,j_3} \gamma^{j_1,j_2,j_3} p^{j_3}_z \frac{\delta}{\delta p_x}
\left[\frac{f^{\hat{n},i_1,i_2,i_3,j_1,j_2,j_3}}{\gamma^{j_1,j_2,\tilde{j_3}}}\right]\Delta V=0,\label{eq:3.5.3}\\
\sum_{j_1,j_2,j_3} \gamma^{j_1,j_2,j_3} p^{j_2}_y \frac{\delta}{\delta p_x}
\left[\frac{f^{\hat{n},i_1,i_2,i_3,j_1,j_2,j_3}}{\gamma^{j_1,\tilde{j_2},j_3}}\right]\Delta V
-\sum_{j_1,j_2,j_3} \gamma^{j_1,j_2,j_3} p^{j_1}_x \frac{\delta}{\delta p_y}
\left[\frac{f^{\hat{n},i_1,i_2,i_3,j_1,j_2,j_3}}{\gamma^{\tilde{j_1},j_2,j_3}}\right]\Delta V=0.\label{eq:3.5.4}
\end{align}
Thus, the energy of particles is described using Eqs.~(\ref{eq:3.5.2})--(\ref{eq:3.5.4}) as follows,
and it is ensured that the energy is not affected by the magnetic field:

\begin{align}
\frac{\delta}{\delta t}\left[ \sum_{j_1,j_2,j_3}\gamma^{j_1,j_2,j_3}mc^2
f^{n+\frac12,i_1,i_2,i_3,j_1,j_2,j_3}\Delta V \right]
+\frac{\delta}{\delta x}\left[ \sum_{j_1,j_2,j_3}\frac{c^2 p_x^{j_1}\gamma^{j_1,j_2,j_3}}{\gamma^{\tilde{j_1},j_2,j_3}}
f^{\hat{n},i_1,i_2,i_3,j_1,j_2,j_3}\Delta V \right]\notag\\
+\frac{\delta}{\delta y}\left[ \sum_{j_1,j_2,j_3}\frac{c^2 p_y^{j_2}\gamma^{j_1,j_2,j_3}}{\gamma^{j_1,\tilde{j_2},j_3}}
f^{\hat{n},i_1,i_2,i_3,j_1,j_2,j_3}\Delta V \right]
+\frac{\delta}{\delta z}\left[ \sum_{j_1,j_2,j_3}\frac{c^2 p_z^{j_3}\gamma^{j_1,j_2,j_3}}{\gamma^{j_1,j_2,\tilde{j_3}}}
f^{\hat{n},i_1,i_2,i_3,j_1,j_2,j_3}\Delta V \right]\notag\\
=\sum_{j_1,j_2,j_3}\left(\frac{qE^{\hat{n},i_1,i_2,i_3}_x p_x^{j_1}}{\gamma^{\tilde{j_1},j_2,j_3}m}
+\frac{qE^{\hat{n},i_1,i_2,i_3}_y p_y^{j_2}}{\gamma^{j_1,\tilde{j_2},j_3}m}
+\frac{qE^{\hat{n},i_1,i_2,i_3}_z p_z^{j_3}}{\gamma^{j_1,j_2,\tilde{j_3}}m}\right)
f^{\hat{n},i_1,i_2,i_3,j_1,j_2,j_3}\Delta V \notag\\
=J^{\hat{n},i_1,i_2,i_3}_x E^{\hat{n},i_1,i_2,i_3}_x+
J^{\hat{n},i_1,i_2,i_3}_y E^{\hat{n},i_1,i_2,i_3}_y+J^{\hat{n},i_1,i_2,i_3}_z E^{\hat{n},i_1,i_2,i_3}_z.\label{eq:3.5.5}
\end{align}
To obtain the energy of the electromagnetic field in discrete form, the temporal product rule Eq.~(\ref{eq:3.1.4}),
and the spatial product rule Eq.~(\ref{eq:3.1.1}) are applied to Eqs.~(\ref{eq:2.5})--(\ref{eq:2.10}):

\begin{align}
\frac{c}{4\pi}E_x^{\hat{n},i_1,i_2,i_3}\left(\frac{1}{c}\frac{\delta}{\delta t}[E_x^{n+\frac12,i_1,i_2,i_3}]
-\frac{\delta}{\delta y}[B_z^{\hat{n},i_1,i_2,i_3}]+\frac{\delta}{\delta z}[B_y^{\hat{n},i_1,i_2,i_3}]\right)
=-J_x^{\hat{n},i_1,i_2,i_3}E_x^{\hat{n},i_1,i_2,i_3},\notag\\
\frac{c}{4\pi}E_y^{\hat{n},i_1,i_2,i_3}\left(\frac{1}{c}\frac{\delta}{\delta t}[E_y^{n+\frac12,i_1,i_2,i_3}]
-\frac{\delta}{\delta z}[B_x^{\hat{n},i_1,i_2,i_3}]+\frac{\delta}{\delta x}[B_z^{\hat{n},i_1,i_2,i_3}]\right)
=-J_y^{\hat{n},i_1,i_2,i_3}E_y^{\hat{n},i_1,i_2,i_3},\notag\\
\frac{c}{4\pi}E_z^{\hat{n},i_1,i_2,i_3}\left(\frac{1}{c}\frac{\delta}{\delta t}[E_z^{n+\frac12,i_1,i_2,i_3}]
-\frac{\delta}{\delta x}[B_y^{\hat{n},i_1,i_2,i_3}]+\frac{\delta}{\delta y}[B_x^{\hat{n},i_1,i_2,i_3}]\right)
=-J_z^{\hat{n},i_1,i_2,i_3}E_z^{\hat{n},i_1,i_2,i_3},\notag\\
\frac{c}{4\pi}B_x^{\hat{n},i_1,i_2,i_3}\left(\frac{1}{c}\frac{\delta}{\delta t}[B_x^{n+\frac12,i_1,i_2,i_3}]
+\frac{\delta}{\delta y}[E_z^{\hat{n},i_1,i_2,i_3}]-\frac{\delta}{\delta z}[E_y^{\hat{n},i_1,i_2,i_3}]\right)=0,\notag\\
\frac{c}{4\pi}B_y^{\hat{n},i_1,i_2,i_3}\left(\frac{1}{c}\frac{\delta}{\delta t}[B_y^{n+\frac12,i_1,i_2,i_3}]
+\frac{\delta}{\delta z}[E_x^{\hat{n},i_1,i_2,i_3}]-\frac{\delta}{\delta x}[E_z^{\hat{n},i_1,i_2,i_3}]\right)=0,\notag\\
\frac{c}{4\pi}B_z^{\hat{n},i_1,i_2,i_3}\left(\frac{1}{c}\frac{\delta}{\delta t}[B_z^{n+\frac12,i_1,i_2,i_3}]
+\frac{\delta}{\delta x}[E_y^{\hat{n},i_1,i_2,i_3}]-\frac{\delta}{\delta y}[E_x^{\hat{n},i_1,i_2,i_3}]\right)=0,\notag\\
\therefore \frac{\delta}{\delta t}\left[\frac{(\mathbf{E}^2+\mathbf{B}^2)^{n+\frac12,i_1,i_2,i_3}}{8\pi}\right]
+\frac{c}{4\pi}\frac{\mathrm{D}}{\mathrm{D}x}[E_y^{\hat{n},i_1,i_2,i_3},B_z^{\hat{n},i_1,i_2,i_3}]
-\frac{c}{4\pi}\frac{\mathrm{D}}{\mathrm{D}x}[E_z^{\hat{n},i_1,i_2,i_3},B_y^{\hat{n},i_1,i_2,i_3}]\notag\\
+\frac{c}{4\pi}\frac{\mathrm{D}}{\mathrm{D}y}[E_z^{\hat{n},i_1,i_2,i_3},B_x^{\hat{n},i_1,i_2,i_3}]
-\frac{c}{4\pi}\frac{\mathrm{D}}{\mathrm{D}y}[E_x^{\hat{n},i_1,i_2,i_3},B_z^{\hat{n},i_1,i_2,i_3}]\notag\\
+\frac{c}{4\pi}\frac{\mathrm{D}}{\mathrm{D}z}[E_x^{\hat{n},i_1,i_2,i_3},B_y^{\hat{n},i_1,i_2,i_3}]
-\frac{c}{4\pi}\frac{\mathrm{D}}{\mathrm{D}z}[E_y^{\hat{n},i_1,i_2,i_3},B_x^{\hat{n},i_1,i_2,i_3}]\notag\\
=-J^{\hat{n},i_1,i_2,i_3}_x E^{\hat{n},i_1,i_2,i_3}_x-
J^{\hat{n},i_1,i_2,i_3}_y E^{\hat{n},i_1,i_2,i_3}_y-J^{\hat{n},i_1,i_2,i_3}_z E^{\hat{n},i_1,i_2,i_3}_z.\label{eq:3.5.11}
\end{align}
Hence, the total energy of the charged particles and electromagnetic field is strictly conserved, even in discrete form,
because the terms on the right-hand-side of Eqs.~(\ref{eq:3.5.5}) and (\ref{eq:3.5.11}) completely cancel out.

\section{Experimental demonstration of conservation property}\label{sec:4}
According to the proposed strategy, a kinetic code is constructed, which is named SPUTNIK:
Structure-Preserving Ultimate Theory as a Numerical Infrastructure for Kinetics.
SPUTNIK is based on the computational theory described in Secs. \ref{sec:2} and \ref{sec:3}.
In this section, code verification is performed via the relativistic two-stream instability
and relativistic Weibel instability.

\subsection{Relativistic two-stream instability}\label{sec:4.1}
In previous studies, non-relativistic two-stream instability \cite{Chen} was calculated as an
electrostatic or Vlasov--Amp\`{e}re test problem.
Here we show the results of relativistic two-stream instability calculated by SPUTNIK in 1D1P mode.
The initial distribution is given by the shifted Maxwell--J\"{u}ttner distribution described as follows:

\begin{align}
f(\mathbf{p})\propto\exp\left(-\alpha\left(\gamma\gamma_0-\frac{\gamma_0\mathbf{v}_0\cdot\mathbf{p}}{mc^2}-1\right)\right),\label{eq:4.1.1}
\end{align}
where $\mathbf{v}_0$ is the velocity of a beam in the observer frame, $\gamma_0=1/\sqrt{1-|\mathbf{v}_0/c|^2}$,
$\alpha=mc^2/k_\mathrm{B}T$, $k_\mathrm{B}$ is the Boltzmann's constant, and $T$ is the temperature.
The velocities of counter-streaming electron beams are $\mathbf{v}_0/c=[\pm 0.8,0,0]^\mathrm{T}$,
and the temperature is $k_\mathrm{B}T=5\mathrm{\ [keV]}$.
The background stationary protons also have a temperature of $k_\mathrm{B}T=5\mathrm{\ [keV]}$.
In this situation, the dispersion relation of a relativistic two-stream instability \cite{Lapenta2007} is described as

\begin{align}
1-\frac{\omega_\mathrm{pi}^2}{\omega^2}
-\frac{\omega_\mathrm{pe}^2}{2\gamma_0^3}\left[
\frac{1}{(\omega-kv_0)^2}+\frac{1}{(\omega+kv_0)^2}\right]=0,\label{eq:4.1.2}
\end{align}
where $\omega$ is the wave frequency, $k$ is the wavenumber, and $\omega_\mathrm{pe}=(4\pi e^2 n_\mathrm{e}/m_\mathrm{e})^{-1/2}$
is the plasma frequency. The imaginary part of $\omega$ corresponds to the growth rate $\Gamma$ of the instability.
Solving Eq.~(\ref{eq:4.1.2}) numerically, the most unstable mode and corresponding growth rate are obtained as follows:

\begin{align}
\frac{kv_0}{\omega_\mathrm{pe}}\simeq 0.28,\quad
\frac{\Gamma}{\omega_\mathrm{pe}}\simeq 0.164.\label{eq:4.1.3}
\end{align}
To calculate the most unstable mode, the length of a periodic domain $L$ is set to be $L\omega_\mathrm{pe}/c=18$,
and the upper/lower limits of the momentum domain are $p/m_\mathrm{i}c=\pm 0.01$ for protons and
$p/m_\mathrm{e}c=\pm 10$ for electrons, respectively.
The number of computational cells is $1024\times1024$.
The temporal interval is given as $c\Delta t/\Delta x=1$.
The implicit method is implemented with the predictor--corrector method, and the number of iterations is 100 per time-step.
A perturbation of wavelength $L$ and amplitude $10^{-5}n_\mathrm{e}$ is given to the electron density.
The initial electric field is set to satisfy Gauss's law, i.e., Eq.~(\ref{eq:3.3.6}).

\begin{figure}
\includegraphics[width=0.95\textwidth]{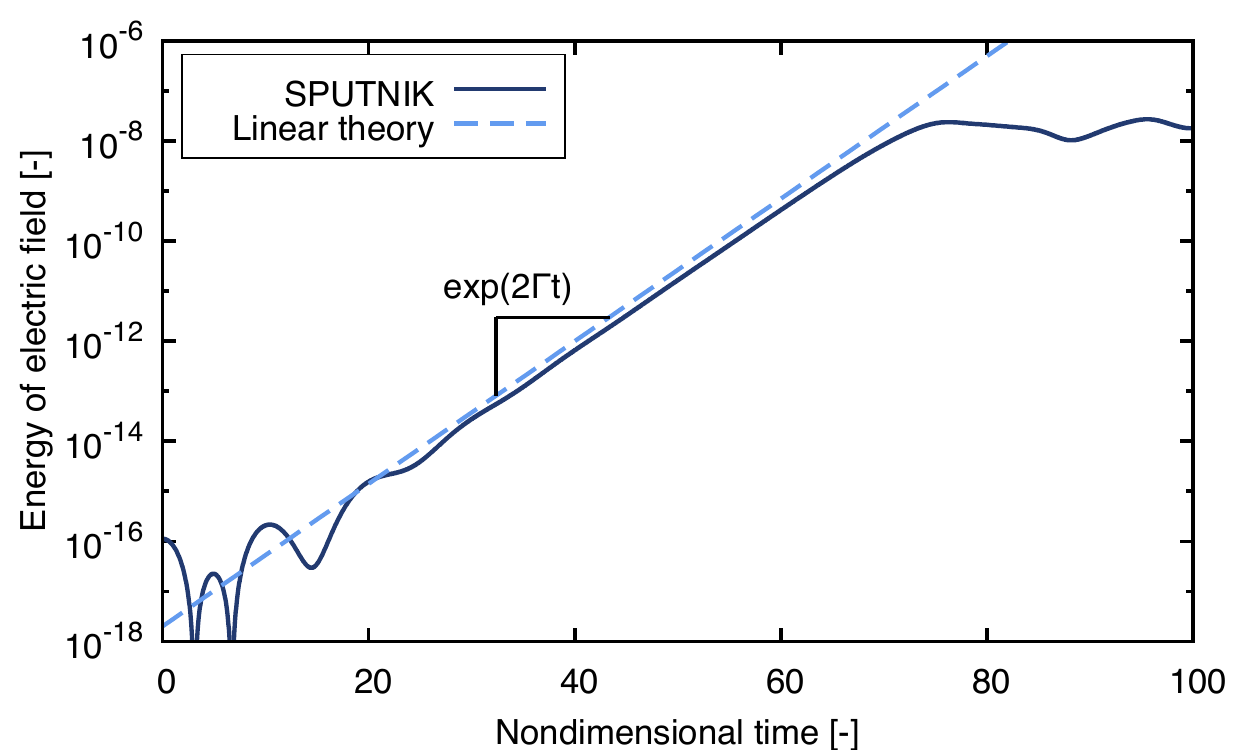}
\centering
\caption{\label{fig:1} (Color online) Amplification of the electric field energy owing to relativistic two-stream instability.
The time is normalized as $\omega_\mathrm{pe}t$.
The growth rate obtained from the linear theory is reproduced by the numerical solution.
}
\end{figure}

\begin{figure}
\includegraphics[width=0.95\textwidth]{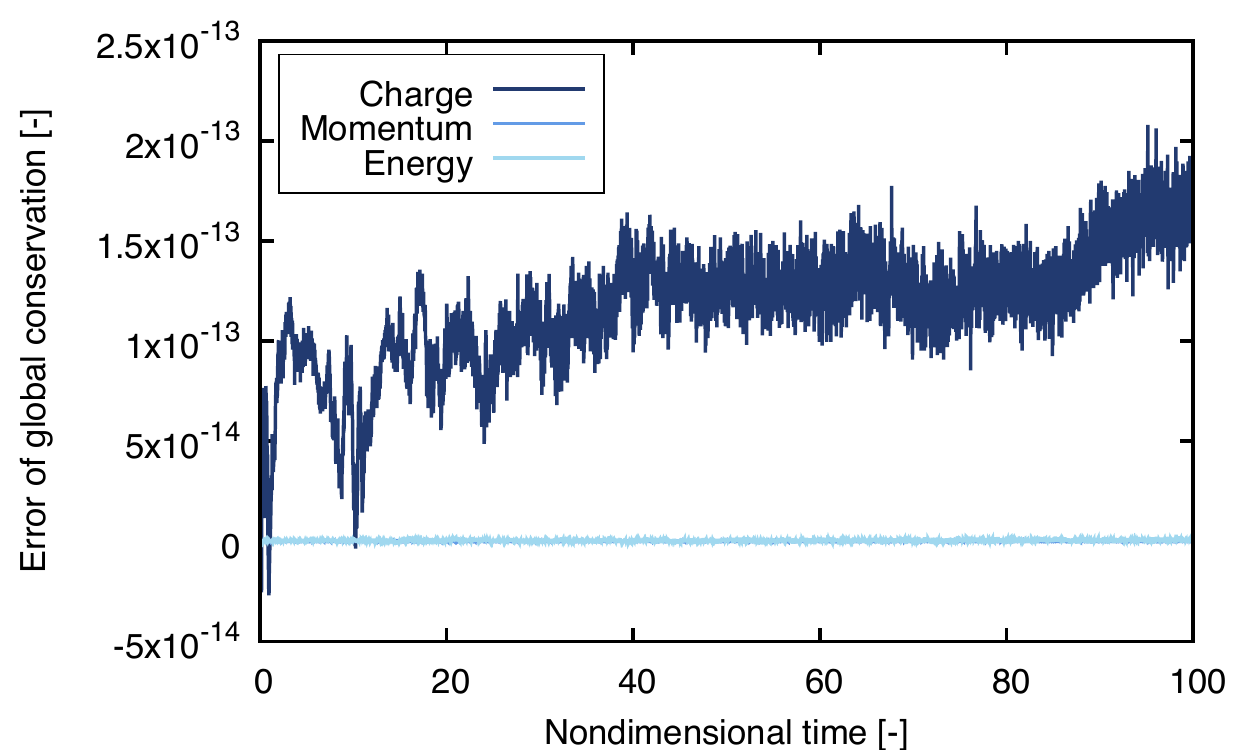}
\centering
\caption{\label{fig:2} (Color online) Conservation property for relativistic two-stream instability solved with the proposed scheme.
The time is normalized as $\omega_\mathrm{pe}t$.
All conservative quantities are strictly preserved only with round-off errors.
}
\end{figure}

\begin{figure}
\includegraphics[width=0.95\textwidth]{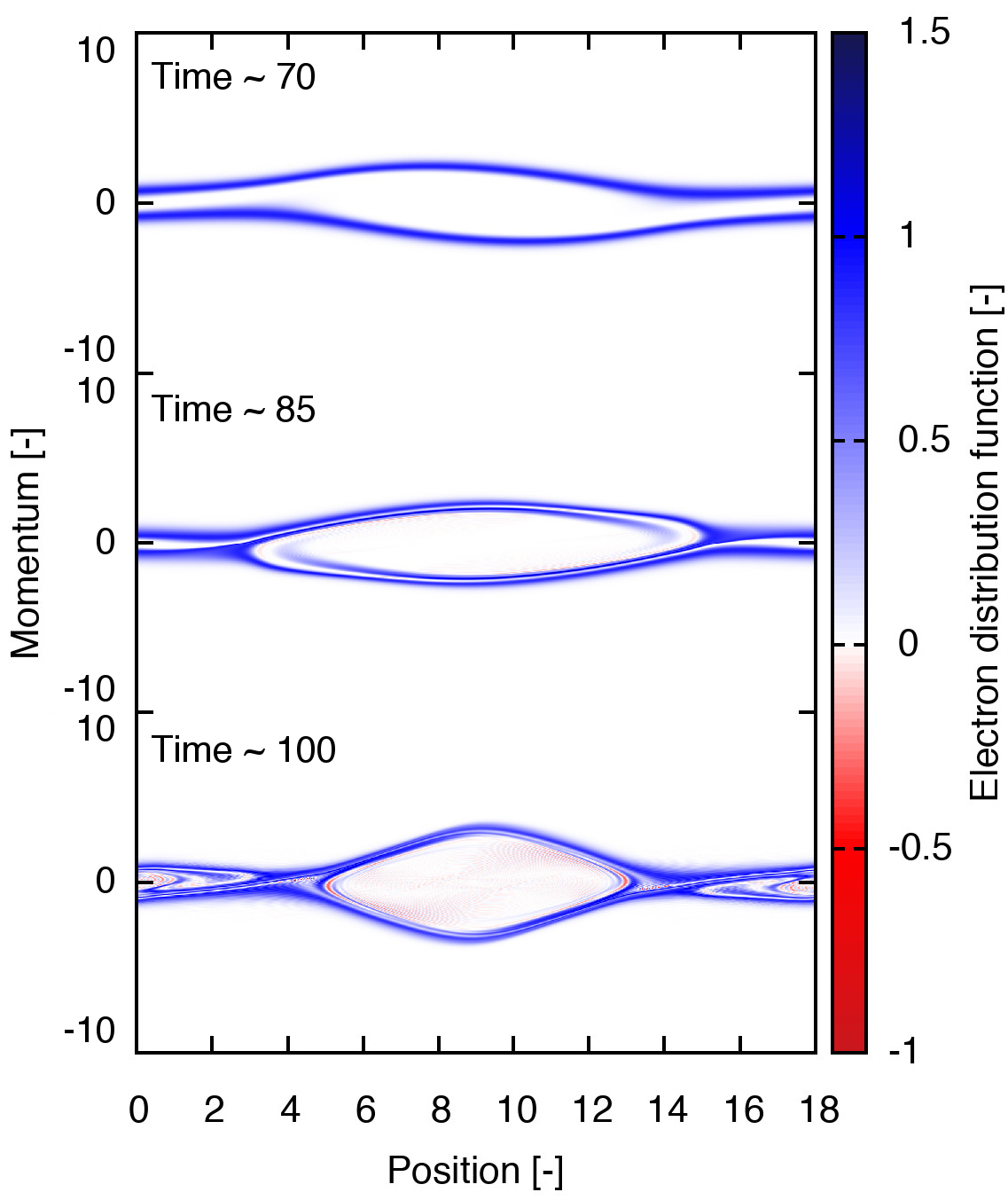}
\centering
\caption{\label{fig:3} (Color online) Electron distribution function of relativistic two-stream instability solved with the proposed scheme.
At a later time, the distribution function becomes negative and the conservation of L1-norm is clearly violated.
}
\end{figure}

Figure \ref{fig:1} shows the time development of the electric field energy.
The time is normalized by the plasma frequency ($\omega_\mathrm{pe}t$).
The energy of the electric field is amplified exponentially at the linear growth phase ($30\le \omega_\mathrm{pe}t\le 70$),
and the numerical growth rate agrees well with the linear theory Eq.~(\ref{eq:4.1.3}).
Subsequently, the amplification of the electric field energy saturates and the instability enters the nonlinear regime.
Figure \ref{fig:2} indicates the errors of global conservation.
As shown in the theoretical proof in Sec. \ref{sec:3}, all errors are strictly maintained at the round-off level,
even if the instability has entered the nonlinear regime.
Thus, the conservation property of the proposed strategy has been demonstrated experimentally.
Note that the proposed scheme cannot maintain the conservation of the L1-norm of $f$ due to the central difference.
We should use many computational cells to mitigate the contamination of the numerical solutions 
by numerical dispersion.
In Fig. \ref{fig:3}, the distribution function of electrons becomes negative at $\omega_\mathrm{pe}t\sim100$;
this is a clear evidence that the proposed scheme cannot maintain the conservation of L1-norm.
Moreover, the central difference does not include any numerical dissipation.
If we employ the upwind difference, for example, the conservation of L1-norm might be maintained even in discrete form.
However, the upwind difference will break up the quadratic conservative scheme since
the mathematical formulae derived in Sec. \ref{sec:3.1} are no longer applicable.
The best strategy to overcome this issue is to extend the proposed scheme to the Vlasov--Fokker--Planck--Maxwell system,
and to introduce physical/artificial collision terms.


\subsection{Relativistic Weibel instability}\label{sec:4.2}
\begin{figure}
\includegraphics[width=0.95\textwidth]{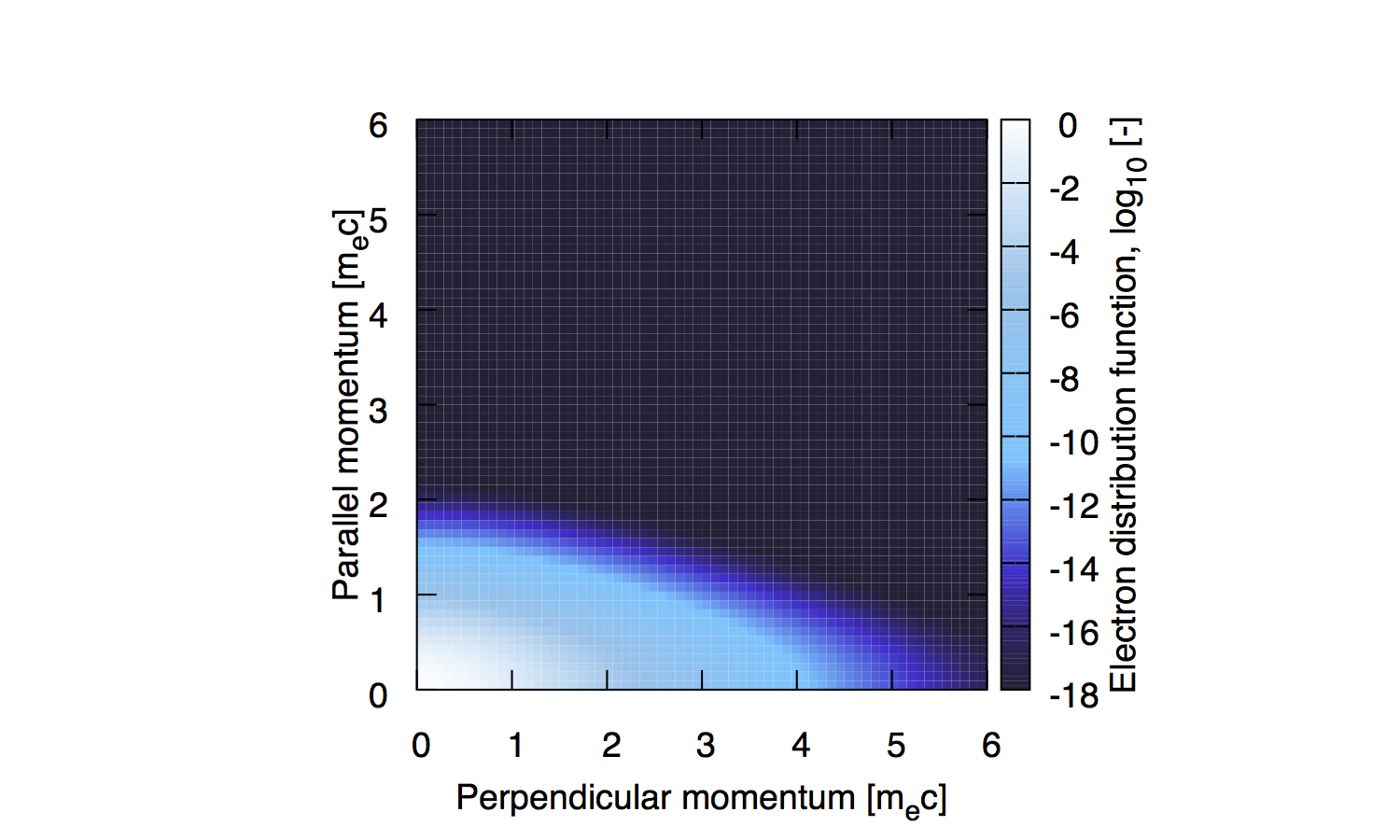}
\centering
\caption{\label{fig:4-0} (Color online) Electron distribution function at the initial state.
}
\end{figure}

In previous studies, the non-relativistic Weibel instability \cite{Weibel1959,Fried1959}
was calculated as an electromagnetic or Vlasov--Maxwell test problem.
Here we show the results of the relativistic Weibel instability calculated by SPUTNIK in the 1D3P mode.
The initial distribution is given by the relativistic bi-Maxwellian distribution described as follows:

\begin{align}
f(\mathbf{p})\propto \exp\left(-\alpha_\perp(\gamma-\gamma_\parallel)-\alpha_\parallel \gamma_\parallel\right),\label{eq:4.2.1}
\end{align}
where $\gamma_\parallel=\sqrt{1+p_\parallel^2/(mc)^2}$,
``$\parallel$'' denotes the parallel direction, i.e., the $x$-direction,
and ``$\perp$'' denotes the perpendicular direction, i.e., the $y$- and $z$-directions.
Temperature anisotropy is given to electrons as $\alpha_\parallel=30$, and $\alpha_\perp=5$,
while for isotropic background protons it is $\alpha_\parallel=\alpha_\perp=5$.
Again, $\alpha=mc^2/k_\mathrm{B}T$ is a reciprocal of the temperature normalized by the rest mass energy.
The electron distribution function at the initial state is shown in Fig.~\ref{fig:4-0}.
In this situation, the dispersion relation for the relativistic Weibel instability \cite{Kaang2009} is given as

\begin{align}
\frac{c^2k^2}{\omega_\mathrm{pe}^2}=\frac{\alpha_\perp^2 \alpha_\parallel}
{\alpha_\parallel K_2(\alpha_\parallel)+\Delta K_1(\alpha_\parallel)}\left[
\frac{\Delta}{\alpha_\perp^2}\left(\frac{K_0(\alpha_\parallel)}{\alpha_\perp}+K_1(\alpha_\parallel)\right)
-\frac{\Gamma}{ck}\int_0^{\infty}\frac{\mathrm{d}\tau}{\xi_\perp^2}\left\{3\Delta\frac{K_0(\alpha_\parallel)}{\xi_\perp^2}
\right.\right.\notag\\
\left.\left.
+\left(3\Delta\frac{\xi_\parallel}{\xi_\perp}+\frac{\xi_\parallel}{\xi_\perp}-\frac{\alpha_\parallel}{\alpha_\perp}\right)
\frac{K_1(\zeta)}{\zeta}+\xi_\parallel^2\frac{\alpha_\parallel}{\alpha_\perp}\frac{K_2(\zeta)}{\zeta^2}\right\}\right],\label{eq:4.2.2}
\end{align}
where $K_n$ is the modified Bessel function of the second kind of order $n$,
$\Delta=\alpha_\parallel/\alpha_\perp-1$, $\xi_\perp=\alpha_\perp+\Gamma\tau/ck$,
$\xi_\parallel=\alpha_\parallel+\Gamma\tau/ck$, and $\zeta=\sqrt{\xi_\parallel^2+\tau^2}$.
By solving Eq. (\ref{eq:4.2.2}) numerically, the most unstable mode and corresponding growth rate are obtained as follows:

\begin{align}
\frac{kc}{\omega_\mathrm{pe}}\simeq 0.92,\quad
\frac{\Gamma}{\omega_\mathrm{pe}}\simeq 0.144.\label{eq:4.2.3}
\end{align}
To calculate the most unstable mode, the length of a periodic domain $L$ is set to be $L\omega_\mathrm{pe}/c=7$,
and the upper/lower limits of the momentum domain are $p/m_\mathrm{i}c=\pm 6$ for protons
and $p/m_\mathrm{e}c=\pm 6$ for electrons, respectively.
The number of computational cells is $128\times128\times128\times128$.
The temporal interval is given as $c\Delta t/ \Delta x=1$.
The implicit method is implemented with the predictor--corrector method,
and the number of iterations is 100 per time-step.
A perturbation with a wavelength is $L$ is given to $B_z$.

\begin{figure}
\includegraphics[width=0.95\textwidth]{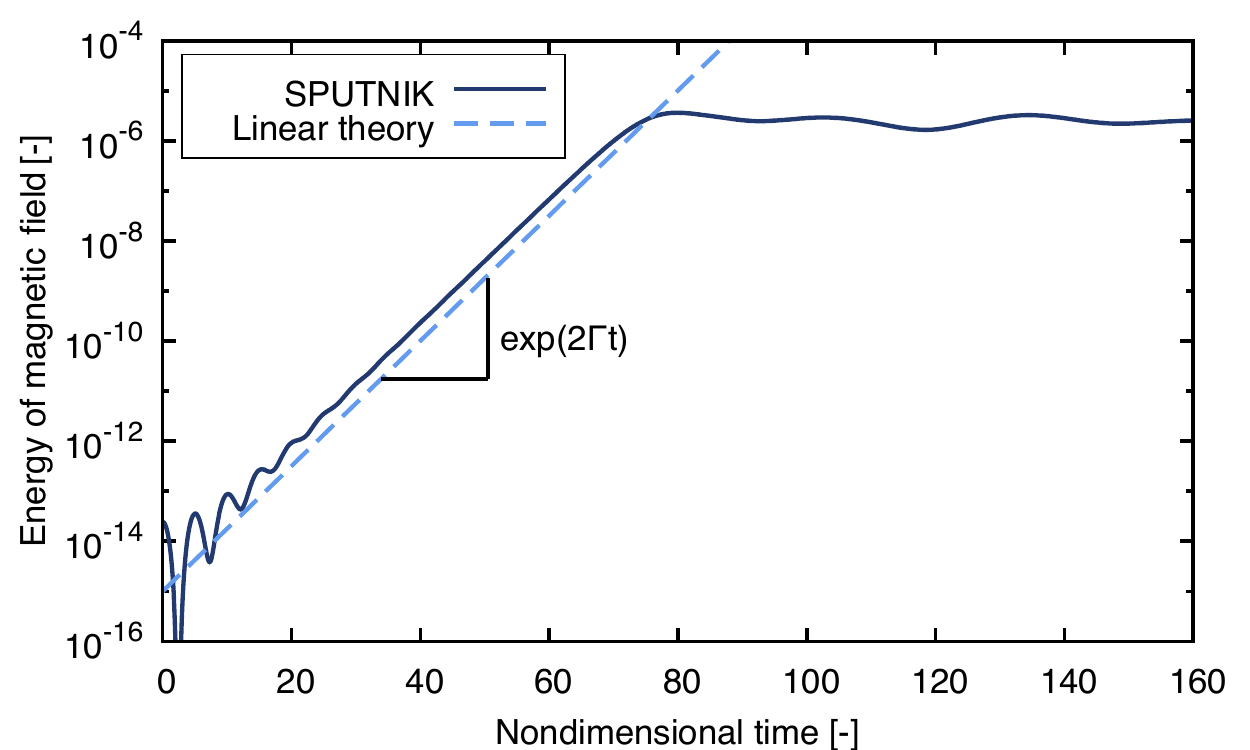}
\centering
\caption{\label{fig:4} (Color online) Amplification of the magnetic field energy owing to relativistic Weibel instability.
The time is normalized as $\omega_\mathrm{pe}t$.
The growth rate obtained from linear theory is reproduced by the numerical solution.
}
\end{figure}

\begin{figure}
\includegraphics[width=0.95\textwidth]{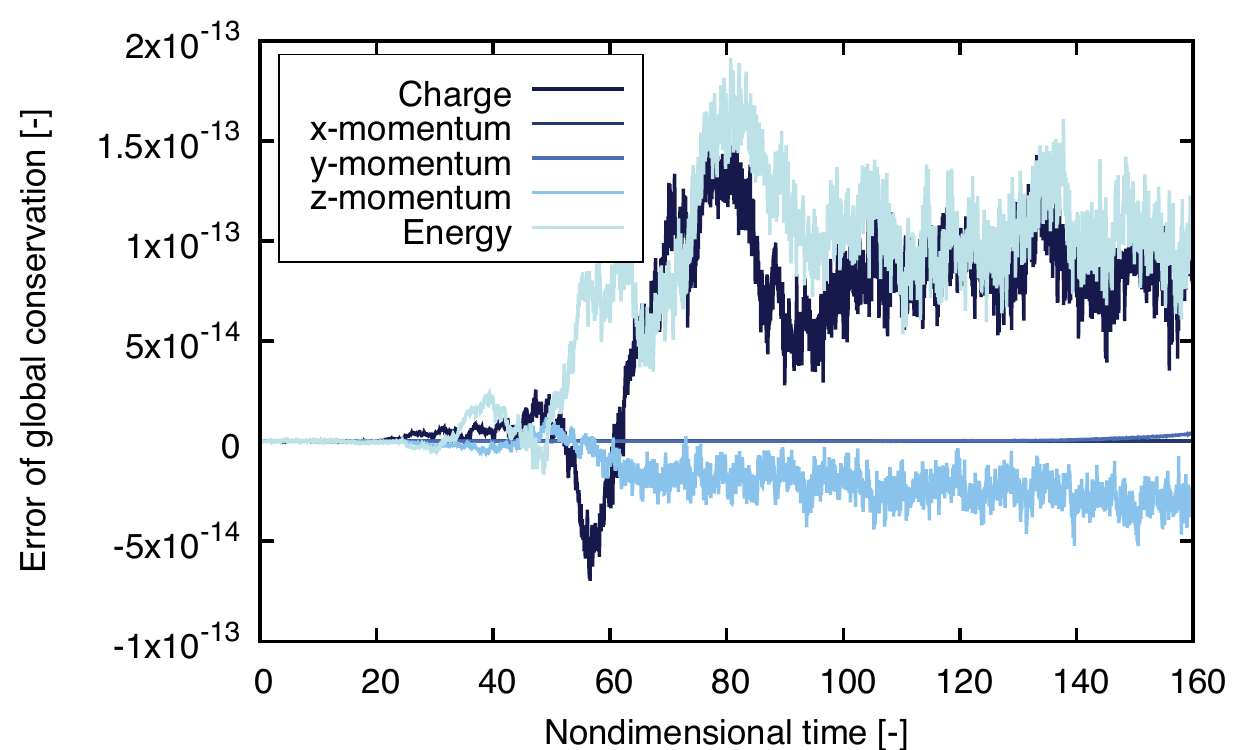}
\centering
\caption{\label{fig:5} (Color online) Conservation property for the relativistic Weibel instability solved with the proposed scheme.
The time is normalized as $\omega_\mathrm{pe}t$.
The conservation laws of charge, momentum, and energy are strictly preserved only with round-off errors.
}
\end{figure}

Figure~\ref{fig:4} shows the time development of the magnetic field energy.
The energy of the magnetic field is amplified exponentially at the linear growth phase ($20\le\omega_\mathrm{pe}t\le70$),
and the numerical growth rate agrees well with the linear theory Eq.~(\ref{eq:4.2.3}).
Subsequently, the amplification of the magnetic field energy saturates and the instability enters the nonlinear regime.
Figure~\ref{fig:5} indicates the errors of global conservation.
As shown in the theoretical proof in Sec.~\ref{sec:3}, all errors are strictly maintained at the round-off level,
even if the instability has entered the nonlinear regime.
Thus, the conservation property of the proposed strategy has been demonstrated experimentally.


\section{Conclusions}\label{sec:5}
In this article, we have presented a quadratic conservative scheme for relativistic Vlasov--Maxwell system
which is composed of a relativistic Vlasov equation, and Maxwell equations.
The scheme is based on the finite-difference method, and thereby enables us to use non-periodic boundary conditions.
We introduced a Crank--Nicolson method for the temporal dimension,
and a central difference for spatial and momentum dimensions.
Before the discretization of the relativistic Vlasov equation, we have to convert it from a non-conservative formulation
to a conservative formulation.
The conservation property of the relativistic Vlasov--Maxwell system can be proven with a product rule, and integration-by-parts.
Accordingly, we have also introduced some mathematical formulae for product rules in discrete form,
and a summation-by-parts.
All terms in the proposed scheme have been carefully designed so that every source term that is generated in the 
momentum and energy equations cancels out even in discrete form.

We constructed a kinetic simulation code named SPUTNIK.
Experimental demonstration of conservation property was performed via
relativistic two-stream instability and relativistic Weibel instability.
In both verification tests, SPUTNIK could maintain errors of
global conservation of charge, momentum, and energy at a round-off level.
It was confirmed that numerical growth rates agree well with
linear stability theories of the instabilities.
In the verification via the relativistic two-stream instability,
an electron distribution function was found to become negative due to a numerical dispersion of the central difference,
and the conservation property of L1-norm was clearly violated.
By applying a quadratic conservative scheme for the relativistic Vlasov--Fokker--Planck--Maxwell system,
this issue could be overcome.
Recently, a mass-, momentum-, and energy-conserving scheme \cite{Taitano2015} for the Rosenbluth--Fokker--Planck
equation \cite{Rosenbluth1957} has been proposed.
To construct the conservative Vlasov--Fokker--Planck--Maxwell scheme,
the development of a fully conservative scheme for the relativistic Landau--Fokker--Planck equation
based on Braams--Karney potential \cite{Braams1987} may be required.

\section*{Acknowledgments}
T.S. wishes to appreciate the fruitful discussions with Dr. Soshi Kawai (Department of Aerospace Engineering, Tohoku University)
and Dr. Shigeru Yonemura (Institute for Fluid Science, Tohoku University).
This work was supported by a Grant-in-Aid for the Japan Society for the Promotion of Science (JSPS) Research Fellows, No. 15J02622.
Numerical experiments were carried out on a vector supercomputer SX-ACE, Cybermedia Center, Osaka University.

\end{document}